 \documentclass[onecolumn,aps,preprint,showpacs,amsmath,amssymb]{revtex4}
\usepackage{graphicx}
\usepackage{latexsym}
\usepackage{xcolor}
 
\usepackage{dcolumn} 
 
\begin{document}
\newcommand{\eq}{\begin{equation}}                                                                         
\newcommand{\eqe}{\end{equation}}             
 
\title{ Self-similar analysis of the time-dependent compressible and incompressible boundary 
layers including heat conduction }

\author{Imre Ferenc Barna$^{1}$, Kriszti\'an Hricz\'o$^{2}$, Gabriella Bogn\'ar$^{2}$,   
and  L\'aszl\'o M\'aty\'as$^{3}$}
\address{ $^1$ Wigner Research Center for Physics, 
\\ Konkoly-Thege Mikl\'os \'ut 29 - 33, 1121 Budapest, Hungary \\
$^2$ University of Miskolc, Miskolc-Egyetemv\'aros 3515, Hungary, \\ 
$^3$Department of Bioengineering, Faculty of Economics, Socio-Human Sciences and 
Engineering, Sapientia Hungarian University of Transylvania,  
Libert\u{a}tii sq. 1, 530104 Miercurea Ciuc, Romania} 
\date{\today}

\date{\today}

\begin{abstract} 
 We investigate the incompressible and compressible heat conducting boundary layer with applying  
the two-dimensional self-similar Ansatz. Analytic solutions can be found for the incompressible case which can be expressed with 
special functions. The parameter dependencies are studied and discussed in details.    
In the last part of our study we present the ordinary differential equation (ODE) system which is obtained for compressible 
boundary layers.   
\end{abstract} 
\maketitle
\section{Introduction}
It is evident that the study of hydrodynamical equations has a crucial role in engineering 
and science as well. It is also clear that numerous classifications exist for 
various flow systems. One class of fluid flows is the field of boundary layer. 
The development  of this scientific field started with the pioneering work of Prandtl 
\cite{prandt} who used scaling arguments and derived that half of the terms of the Naiver-Stokes 
 equations are negligible in boundary layer flows. 
In 1908 Blasius \cite{blasius} gave the solutions of the steady-state incompressible two-dimensional laminar 
boundary layer equation forms on a semi-infinite plate which is held parallel to a constant 
unidirectional flow.  Later Falkner and Skan \cite{falkner,falkner1} generalized the solutions for 
steady two-dimensional laminar boundary layer that forms on a wedge, i.e. flows in which the plate is 
not parallel to the flow. 
An exhaustive description of the hydrodynamics of boundary layers can be found in the 
classical textbook of Schlichting \cite{sch} recent applications in engineering is discussed by Hori 
\cite{hori}.  
The mathematical properties of the corresponding partial differential equations (PDEs) attracted 
remarkable interest as well. Without completeness we mention some of the available mathematical  
results. Libby and Fox \cite{libby} derived some solutions using perturbation method.    
Ma and Hui \cite{ma} gave similarity solution to the boundary layer problems.  
Burde \cite{burde1,burde2,burde3} gave additional numerous explicit analytic solutions in the nineties. 
Weidman \cite{weid} presented solutions for boundary layers with additional cross flows. 
Ludlow and coworkers \cite{lud} evaluated and analyzed solutions with similarity methods as well. 
Vereshchagina \cite{ver} investigated the spatial unsteady boundary layer equations with group fibering. 
Polyanin in his papers \cite{poy1,poy2} presents numerous independent solutions derived with 
various methods like general variable separation. Makinde \cite{mak} investigated the laminar falling liquid film with
variable viscosity along an inclined heated plate problem using perturbation technique together with a
special type of Hermite – Pad\'e  approximation. In nanofluids the importance of buoyancy \cite{AnSa2016}, aspects on 
bioconvection \cite{MaAn2016}, and possible modified viscosity \cite{SaKoAn2016} are also discussed. 
One may find exact solutions for the oscillatory shear flow in \cite{SaGiKo2017,SaGi2018}.

Bogn\'ar \cite{bogn} applied the steady-state boundary layer flow equations for non-Newtonian fluids and presented 
self-similar results. Later it was generalized \cite{bognhri},  and the steady-state heat conduction mechanism was included in the calculations as well.   
Certain parameters of the nanofluid can be tuned by varying the amount of nanoparticles in the fluid \cite{MaEbTe1993, Ch1995,Ng2007}.

In our former studies we investigated three different kind of Rayleigh-B\'enard heat conduction problems \cite{imre1,imre2,imre3} 
which are full two-dimensional viscous flows coupled to the heat conduction equation.  
We might say that the heated boundary layer equations - from the mathematical point of view - show some similarities to the Rayleigh-B\'enard problem. 
These last five publications of us \cite{bogn,bognhri,imre1,imre2,imre3} led us to the decision that it would be worst examining  
 heated boundary layers with the self-similar Ansatz. 
 
In the following  we apply the Sedov type self-simiar Ansatz \cite{sedov,zeldovich} to the original partial differential equation (PDE) systems 
 of incompressible and compressible boundary layers with heat conduction and reduce them to coupled non-linear ordinary differential equation (ODE) system. For the incompressible case the ODE system  can be solved with quadrature giving analytic solutions for the velocity, pressure and temperature fields. Due, to our knowledge there are no self-similar solutions known and analyzed for any type of time-dependent boundary layer equations including heat conduction. 
\section{Theory}
\subsection{The incompressible case }
 We start with the PDE system of  
\begin{eqnarray}
\frac{\partial u}{\partial x} + \frac{\partial v}{\partial y} &=& 0,  \\ 
\frac{\partial p}{\partial  y} &=&0, \\ 
\rho_{\infty}  \frac{\partial u}{\partial  t} + \rho_{\infty} \left(   u\frac{\partial u}{\partial x} + v\frac{\partial u}{\partial y} 
\right) &=& \mu \frac{\partial^2 u}{\partial  y^2} 
- \frac{\partial p}{\partial  x}, \\ 
 \rho_{\infty}c_p   \frac{\partial T}{\partial t} +  \rho_{\infty}c_p \left( u  \frac{\partial T}{\partial x}  + v \frac{\partial T}{\partial y} 
\right)  &=&  \kappa  \frac{\partial^2 T}{\partial  y^2}, 
\label{pde} 
\end{eqnarray}
where the dynamical variables are the two velocities components $u(x,y,t), v(x,y,t)$ of the fluid  the pressure $p(x,y,t)$ and the temperature $T(x,y,t)$.  
The additional physical parameters are  $\rho_{\infty}, c_p, \mu, \kappa,$  
the fluid density at asymptotic distances and times, the heat capacity at fixed pressure, 
the kinematic viscosity and the thermal diffusivity, respectively.  
It is important to emphasize at this point, that this description for the heated boundary layer is valid for small 
velocities in laminar flow, only.  More information can be found in the classical book of Schlichting \cite{sch}  ($8^{th}$ addition page 211).  Outside the laminar flow regime a viscous heating term should be added to the final temperature equation with the form of $ \mu (u_y)^2$.  (A similar  analysis for that system is already in progress and will be the topic of our next distinct study.)   

There is no general fundamental theory for nonlinear PDEs, but over time, some intuitive methods have evolved, most of them can be derived from symmetry considerations. Numerous  (almost arbitrary) functions can be constructed which couple the temporal and spatial variables to a new reduced variable from intuitive reasons. Our long term experience shows that two of them are superior to all others and have 
direct physical meanings. These are the traveling wave and the self-similar Ans\"atze. The first is more or less well known from the 
community of physicists and engineers and, has the form of $G(x,t) = f(x \mp ct) $ and we may call $\eta = x \mp ct$ as the new reduced variable,  where c is the propagation speed of the corresponding wave. Here $G(x,t)$ is the investigated dynamical variable in the PDE.  
$G(x,t)$ could be any physically relevant property, like temperature, electric field or the like. 
This Ansatz can be applied to any kind of PDE and will mimic the general wave property of the investigated physical system. 

The second (and not so well known) is the self-similar Ansatz with the from of $ G(\eta) =  t^{-\alpha} f(x/t^{\beta})$. 
There $\alpha$ and $\beta$ are two free real parameters, it can be shown that this Ansatz automatically gives the Gaussian or fundamental solution of the diffusion (or heat conduction) equation. In general, and this is the key point here, this trial functions helps us to get 
a deeper insight into the dispersive and decaying behavior of the investigated physical system. This is the main reason why we use it  
in this form. Viscous fluid dynamic equations automatically fulfill this condition, therefore it is highly probable, that this Ansatz 
leads to physically rational solutions.    
It is easy to modify the original form of the Ansatz to two (or even three) spacial dimensions and generalize it to multiple dynamical variables, hereupon we apply the following form of:   
\begin{eqnarray}
u(x,y,t) &=& t^{-\alpha} f(\eta), \hspace*{1cm}
v(x,y,t)  = t^{-\delta} g(\eta),  \nonumber \\    
T(x,y,t) &=&  t^{-\gamma}h(\eta), \hspace*{1cm}
p(x,y,t) = t^{-\epsilon} i(\eta), 
\label{ansatz}
\end{eqnarray}
with the new argument $\eta = \frac{x+y}{t^{\beta}}$ of the shape functions. 
(To avoid later physical interpretation problems of negative values we define temperature as a temperature difference relative to the average $ T = \tilde{T} - T_{av} $.) 
All the exponents $\alpha,\beta,\gamma,\delta $ are real numbers. (Solutions with integer exponents are called 
self-similar solutions of the first kind, non-integer exponents generate self-similar solutions of the second kind.)  
It is important to emphasize that the obtained results fulfill well-defined initial and boundary problems of the original PDE system via fixing their integration constants of the derived ODE system. 

The shape functions $f,g,h$ and $i$ could be any continuous functions with existing first and second continuous derivatives  and will be evaluated later on.  
The logic, the physical and geometrical interpretation of the Ansatz were exhaustively analyzed in 
all our former publications \cite{imre1,imre2,imre3} therefore we skip it here. 
 The general scheme of the calculation, how the self-similar exponents can derived is given in \cite{imre4} in details. 
The main idea is the following: after having done the spatial and temporal derivatives of the Ansatz  the obtained terms should be 
replaced into the original PDE system. Due to the derivations all terms pick up an extra time dependent factor like $t^{-\alpha-1}$  
or $t^{-2\beta}$ because of the reduction mechanism the new variable of the shape functions is now  $\eta$ therefore all kind 
of extra time dependences have to be canceled. Therefore all the exponents of the time dependences eg. $\alpha +1$ or $2\beta$ should 
cancel each other which dictates a relation among the self-similar variables. In our very first paper we gave all the details of this kind of a calculation for the non-compressible newtonian three dimensional Navier-Stokes equation \cite{imre4}.   

The main points are, that $ \alpha, \delta, \gamma, \epsilon$ are responsible for the rate 
of decay and $\beta$ is for the rate of spreading of the corresponding dynamical variable for positive exponents. 
 Negative self-similar exponents (except for some extreme cases) 
mean unphysical, exploding and contracting solutions.     
The numerical values of the exponents are now the following:  
\eq
\alpha = \beta = \delta = 1/2, \hspace*{1cm} \epsilon = 1, \hspace{1cm}  \gamma = \textrm{arbitrary real number}.   
\eqe 
Exponents with numerical values of one half mean the regular Fourier heat conduction 
(or Fick's diffusion) process. 
One half values for the exponent of the velocity components and unit value exponent for the pressure decay are usual for the incompressible Navier-Stokes equation \cite{imre4}. 

The obtained ODE system reads 
\begin{eqnarray}
f' + g' &=&  0, \\ 
i' &=& 0 , \\    
\rho_{\infty} \left(- \frac{f}{2} - \frac{f' \eta}{2} \right) + \rho_{\infty}(ff' +gf')& =& \mu f'' - i', \\ 
\rho_{\infty}c_p \left(- \gamma h - \frac{h'\eta}{2} \right) +  \rho_{\infty}c_p(fh' + gh') &=& \kappa h'',
\label{ode1}
\end{eqnarray}
where prime means derivation in respect to the variable $\eta$. 
The first two equations are total derivatives and can be integrated directly yielding:  $f + g = c_1$ and $i = c_2$. 
Having total derivatives in a dynamical systems automatically mean conserved quantities, (the first of them is now mass conservation).  
After some straightforward algebraic manipulation we arrive to a separate second order ODE for the velocity shape which is also 
a total derivative and can be integrated leading to:  
\eq
 \mu f' +\rho_{\infty}f \left(\frac{\eta}{2} - c_1\right) - c_2= 0, 
\eqe
with the analytic solution of 
 \begin{eqnarray}
f=&   
\left( \frac{c_2 \sqrt{\pi} e^{-\frac{\rho_{\infty} c_1^2}{\mu}} \cdot   \emph{erf}\left[  \frac{1}{2} \sqrt{-\frac{\rho_{\infty}}{\mu}}\eta + \frac{\rho_{\infty c_1}}{\sqrt{-\mu \rho_{\infty} }}   \right]    }{\sqrt{-\mu \rho_{\infty}}} + c3 \right) \cdot
   \emph{e}^{\frac{\eta(-\eta + 4c_1)\rho_{\infty}}{4\mu}}  
\label{f_solu}
\end{eqnarray}  
where erf means the usual error function \cite{NIST}. Note, that for the positive 
real constants $\rho_{\infty}, \mu$ the complex 
quantity $\sqrt{-\rho_{\infty} \mu}$  appears in the argument of the error functions 
and as a complex multiplicative prefactor simultaneously making the final result a pure real function. The second important thing is to note, that for 
the $c_1 = c_2 = 0 $ trivial integration constants the solution is simplified to the Gaussian function of 
\eq
f = c_4 e^{ -\frac{\rho_{\infty \eta^2}}{4\mu}}. 
\eqe
This means that the velocity flow process shows similarity to the regular diffusion of heat conduction phenomena. 
Similar solutions (containing exponential and error functions) were found for the stationary velocity field by Weyburne in 2006 with probability distribution function
methodology \cite{wey}. 

Figure (1) shows the  general velocity shape function (\ref{f_solu}) for various parameter sets.    
The choice of these parameters 
are arbitrary, we are not limited to real fluid parameters,  however we try to create the most general and most informative figures, 
which mimic the general features of the solution function. 
The functions are the modification of the error function. The crucial parameter is the  ratio $\rho_{\infty}/\mu$, 
if this is larger than unity then the function tends to a sharp Gaussian. 
\begin{figure} 
\scalebox{0.9}{
\rotatebox{0}{\includegraphics{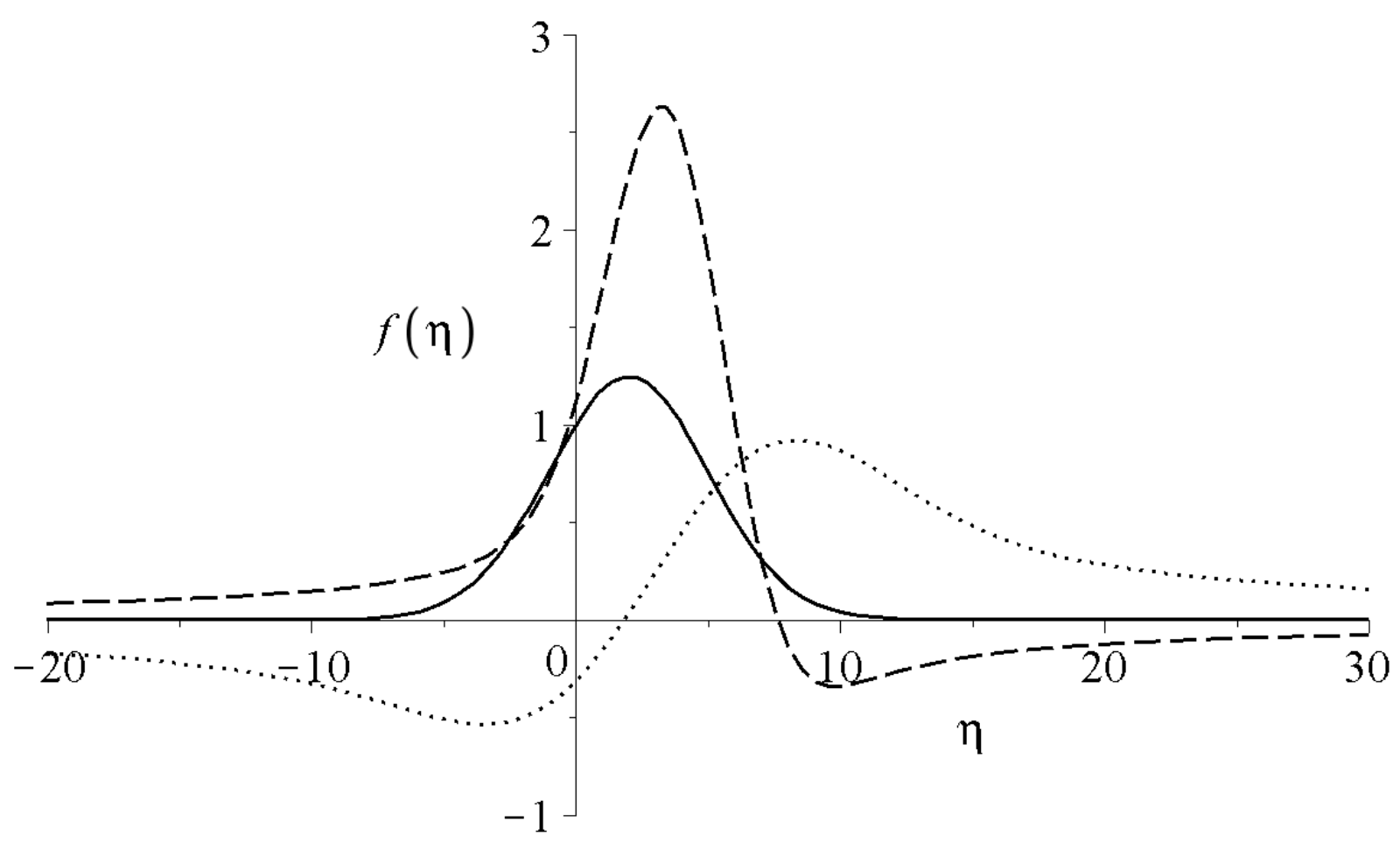}}}
\vspace*{-0.5cm}
\caption{The graphs of the velocity shape function $f(\eta)$  in Eq. (\ref{f_solu}) for three different parameter sets 
($c_{1},c_{2},c_{3},\mu,\rho_{\infty}$). 
The solid, dashed and dotted lines are for $(1, 0, 1, 4.1, 0.9 )$, 
 $(2, -1, 0.5, 2.5, 1)$ and $(2, 2, 0.3, 10, 1)$, respectively.}
\label{egyes}    
\end{figure} 
\begin{figure} 
\scalebox{0.9}{
\rotatebox{0}{\includegraphics{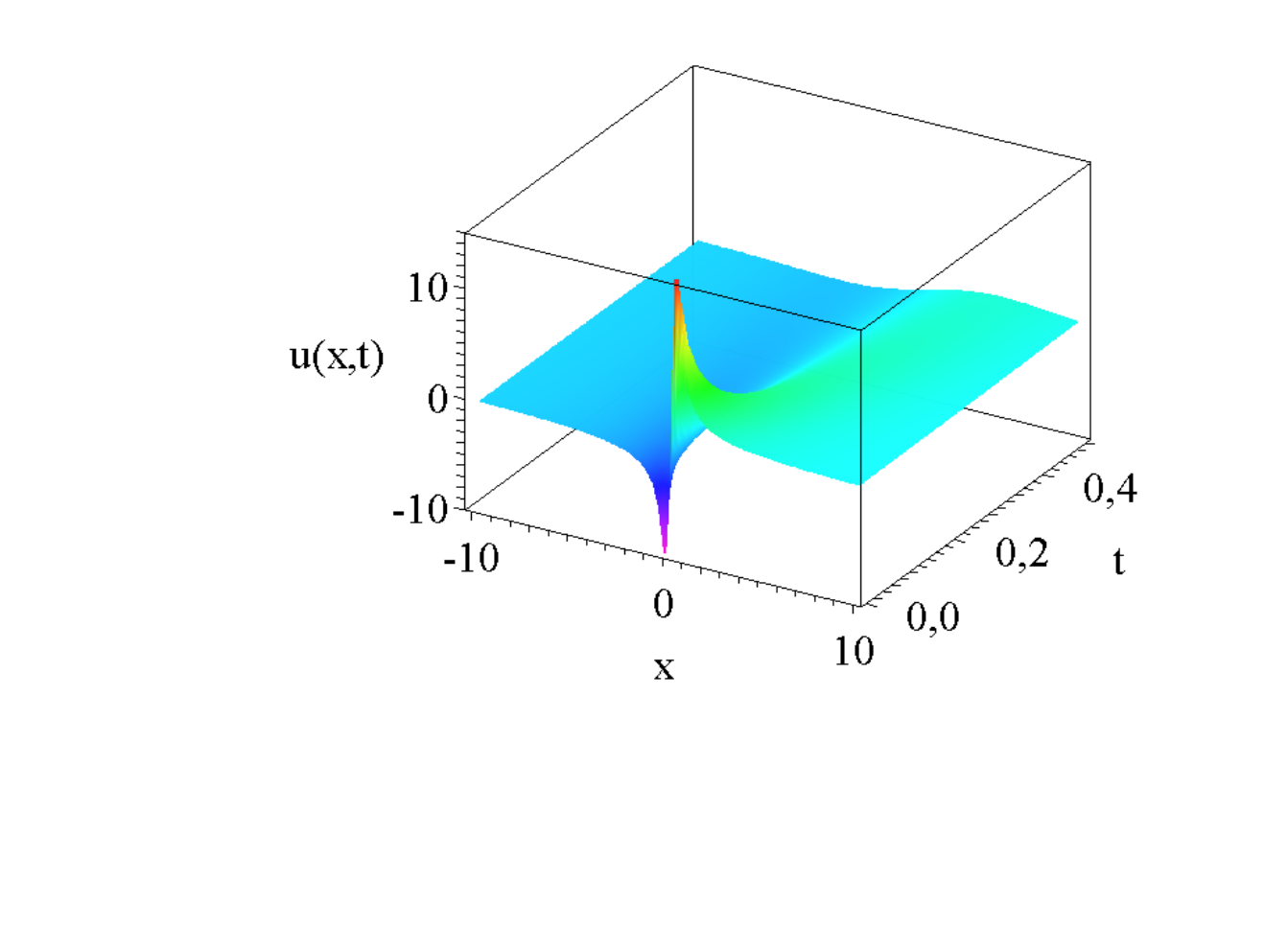}}}
\vspace*{-2.5cm}
\caption{The velocity distribution function $u(x,y=0,t) = \frac{1}{t^{1/2}}f(\eta) $ for the 
third parameter set presented on the previous figure.}
\label{kettes}    
\end{figure}
Figure (\ref{egyes}) presents the velocity distribution function. Note, the very sharp peak in the origin and 
the extreme quick time decay along the time axis.   

There is a separate ODE for the temperature distribution as well  
\eq
\frac{\kappa}{\rho_{\infty} c_p} h'' - h'\left(c_1 - \frac{\eta}{2} \right)  + \gamma h = 0. 
\label{temp}
\eqe
 For the most general case (when $\gamma$ is an arbitrary real number,) and  $ c_1 \ne 0$  the solutions of Eq. (\ref{temp}) can 
be expressed with the Kummer M and Kummer U functions \cite{NIST}  
\eq
h = c_2 M\left(\gamma, \frac{1}{2}; 	-\frac{c_p \rho_{\infty} [\eta - 2c_1]^2}{4\kappa}  \right) + c_3 U\left(\gamma, \frac{1}{2};  -\frac{c_p \rho_{\infty} [\eta - 2c_1]^2}{4\kappa}  \right). 
\label{h_solu}
\eqe 
M is regular in the origin and U is irregular, therefore we investigate only the 
properties of M which means $(c_3 =0)$. 
The M and U functions form a complete orthogonal function system if the argument is linear. Now, the 
argument is quadratic, - in our former studies we found numerous such solutions, for incompressible  \cite{imre4} or for 
compressible \cite{imre5} multidimensional Navier-Stokes or Euler 
equations -- however, we still do not know the physical message of this property. 

It can be easily proven with the definition of the Kummer functions using the Pochhammer symbols \cite{NIST}, that for negative integer $\gamma$ values our results can be expanded into finite order polynomials, which are divergent for large arguments $\eta$. For non-integer $\gamma < 0$ values, we get infinite divergent polynomials as well. 

\begin{figure} 
\scalebox{0.8}{
\rotatebox{0}{\includegraphics{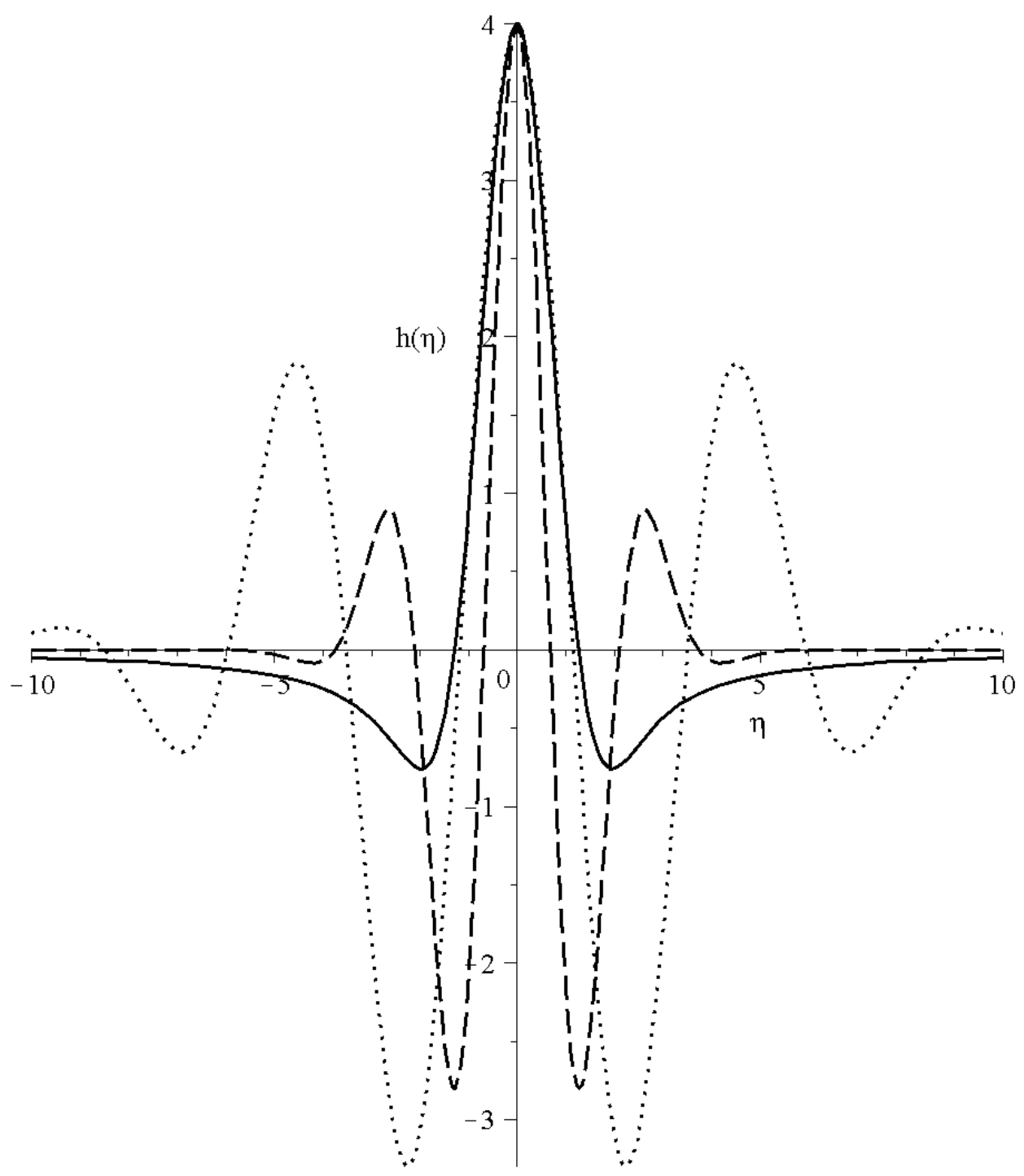}}}
\caption{The graphs of the temperature shape function Eq. (\ref{h_solu}) 
for three different parameter sets ($\gamma, c_2,c_3,c_p,\rho_{\infty}, \kappa$). 
The solid, dashed and dotted lines are for $(0.8, 4, 0, 1, 0.9, 0.3 )$, 
 $(3.4, 4, 0, 1, 1, 0.6)$ and $(6.3, 4, 0, 1, 3, 10)$, respectively.}
\label{harmas}    
\end{figure}

\begin{figure} 
\scalebox{0.8}{\hspace*{-1cm}
\rotatebox{0}{\includegraphics{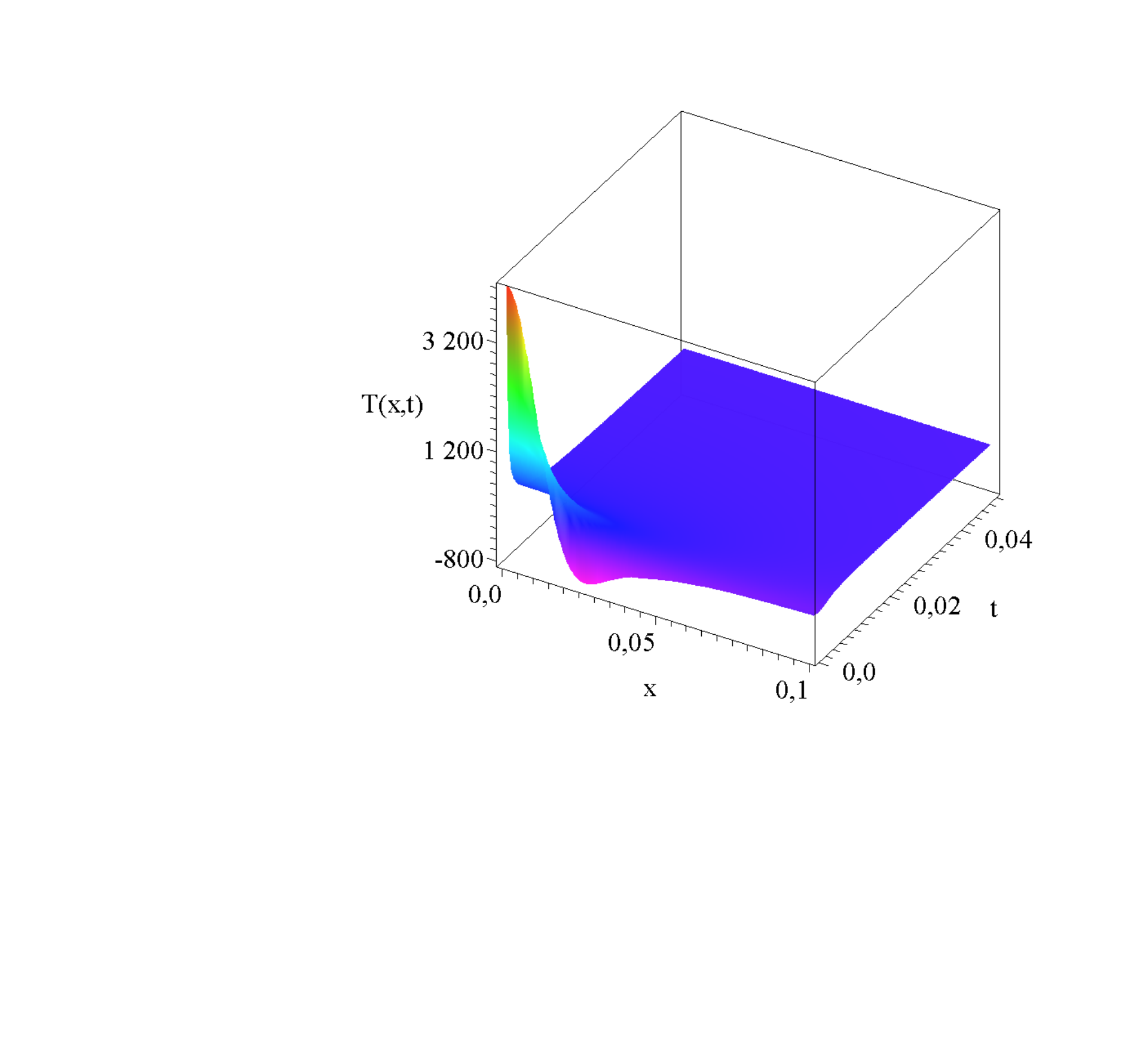}}}
\vspace*{-3.5cm}
\caption{The temperature distribution function $T(x,y=0,t) = \frac{1}{t^1}h(\eta) $ for the first parameter set presented on the previous figure.}
\label{negyes}    
\end{figure}

The most relevant parameter of the solutions is evidently $\gamma$. 
The integral constant $c_1$ just shifts the solutions parallel to the $x$ axis, $c_2$ scales the solutions, and $c_p \rho_{\infty}/ \kappa$ parameter 
just scales the width of the solution.  
Figure (\ref{harmas}) presents three different solutions for various positive $\gamma$ values. (All negative $\gamma$ values mean divergent shape functions for large $\eta$s which are unphysical and outside of our scope.)  
Note, larger $\gamma$s mean more oscillations. 
For a better understanding we present the projection of the total solution of the temperature field $T(t,x,y)$ on Figure (\ref{negyes}) for the $y = 0$ coordinates.   

For some special values of $\gamma$ the temperature shape function can be expressed with other simpler special functions. For values of $\gamma = \pm \frac{1}{2}$ and $ 0$ the shape functions all contain the error function. Negative integer $\gamma$s result even order polynomials.  (E.g. $\gamma = -1$ defines the shape function of 
$f = (c_2 + c_3)\cdot(2\kappa + c_p \rho_{\infty}[\eta-2c_1]^2)$. ) 
Polynomials are divergent in infinity therefore are out of our physical interest.    
 
For the sake of completeness we present the solutions for the pressure as well. 
The ODE of the shape function is trivial with the solution of: 
\eq
i' = 0, \hspace*{1cm} i = c_4.
\eqe
Therefore, the final pressure distribution reads: 
\eq
p(x,y,t) = t^{-\epsilon} \cdot i(x,y,t) = \frac{c_4}{t}, 
\eqe
which means that the pressure is constant in the entire space at a given time point, but has a quicker time decay than the velocity field. 

\subsection{The compressible case}
In the last part of our study we investigate the compressible boundary layer equations. The starting PDE system is now changed to the following: 
\begin{eqnarray}
 \frac{\partial \rho}{\partial t}   +  \frac{\partial \rho}{\partial x} u +
\rho \frac{\partial u}{\partial x}  +  \frac{\partial \rho}{\partial y} v +
\rho \frac{\partial v}{\partial y}  
 &=& 0,  \\ 
\frac{\partial p}{\partial  y} &=&0, \\ 
\rho   \frac{\partial u}{\partial  t} + \rho \left(   u\frac{\partial u}{\partial x} + v\frac{\partial u}{\partial y} 
\right) &=& \mu \frac{\partial^2 u}{\partial  y^2} 
- \frac{\partial p}{\partial  x}, \\ 
  c_p  \rho  \frac{\partial T}{\partial t} +  c_p \rho \left( u  \frac{\partial T}{\partial x}  + v \frac{\partial T}{\partial y} 
\right)  &=&  k  \frac{\partial^2 T}{\partial  y^2}, 
\label{pde} 
\end{eqnarray}
the notation of all the variables are the same as for the incompressible case. For closing constitutive equation (or with other name "equation of state" (EOS)) we apply the ideal gas $ p = R\rho T$ where $R$ is the universal gas constant. 
(Of course, there are numerous EOS available for physically relevant materials, and each gives us an additional 
new system to investigate, but that lies outside the scope of our present study.)
For the dynamical variables we apply the next self-similar Ansatz of:  
 \begin{eqnarray}
\rho(x,y,t) &=& t^{-\alpha} f(\eta), \hspace*{1cm}  
u(x,y,t) = t^{-\gamma} g(\eta), \\    
v(x,y,t) &=& t^{-\delta}h(\eta),  \hspace*{1cm}
T(x,y,t) = t^{-\epsilon} i(\eta),  
\label{ansatz2}
\end{eqnarray}
with the usual new variable of $\eta = \frac{x+y}{t^{\beta}}$.  

To obtain a closed ODE system the following relations must held for the similarity exponents 
\eq 
\alpha = 0,  \hspace{1cm}  \beta = \delta = \gamma = \epsilon = 1/2.  
\eqe 
Note, that now all the exponents have fixed numerical values. 
The $\alpha = 0$ means two things, first the density as dynamical variable has no spreading property (just decay $\beta > 0$), second, the first continuity ODE is not a total derivative and cannot be integrated directly. 
This system has an interesting peculiarity, our experience showed, that the  incompressible Navier-Stokes (NS) equation \cite{imre4} has all fixed self-similar 
exponents and the compressible one \cite{imre5} has one free exponent. 
It is obvious that an extra free exponent makes the mathematical structure richer leaving more room to additional solutions.     
(As we mentioned above, self-similar exponents with the value of one half has a close 
connection to regular Fourier type heat conduction mechanism.)  
Parallel, the obtained ODE system reads 
\begin{eqnarray}
-\frac{1}{2}\eta f' + fg' +f'g + f'h + fh' &=&  0,  \label{comp1} \\ 
R(f'i + fi') &=& 0 ,  \label{comp2}\\    
 f \left(- \frac{g}{2} - \frac{g' \eta}{2} \right) + f(gg' +gh')& =& \mu g'' - R(f'i + fi'), \label{comp3} \\ 
c_p f \left( -\frac{i}{2}  - \frac{i'\eta}{2} \right) +  c_p f (gi' + hi') &=& \kappa i'', \label{comp4}
\label{ode2}
\end{eqnarray}
where prime means derivation in respect to $\eta$.

Having done some non-trivial algebraic steps a decoupled ODE can be derived for the 
density field. First, the pressure equation (\ref{comp2}) can be integrated, then $i(\eta)$ can be expressed, 
after the derivatives $i'$ and $i''$ can be evaluated, then plugging it into (\ref{comp4}) the $(g+h)$ quantity can be expressed with $f, f'$ and $f''$.
Finally, calculating the derivatives of  $(f+g)$ and substituting them into (\ref{comp1}) an independent ODE can be deduced for the density shape function. 
These algebraic manipulations are more compound and contain many more steps what we had in the past for various flow systems like  \cite{imre3,imre4}. 
With the conditions $f(\eta) \ne 0 $ and $f'(\eta) \ne 0$, the next highly non-linear ODE can be derived 
\eq
-\kappa f' f^2 f''' + f''\left(\kappa f^2 f'' + 2\kappa f f'^2 +  \frac{1}{2} c_p f^4 \right) + f'^2 \left(-2\kappa f'^2 -  c_p f' f^2 \cdot
\eta - \frac{3}{2}c_p f^3 \right ) = 0.  
\label{f_dens_compr}
\eqe 
Such ODEs have no analytic solutions for any kind of parameter set (of course $ \kappa \ne 0$ and 
$c_p \ne 0$).  Therefore, pure numerical integration processes have to be applied. 
We have to mention, that  an analogous fourth-order non-linear ODE was derived in the viscous heated B\'enard system 
\cite{imre3} and was analyzed with numerical means.     

The shape function of the temperature field can be easily derived from (\ref{comp2}) without any 
additional derivation 
\eq
i = \frac{c_1}{R f}.
\label{temp_shape_compr}
\eqe
We have to note two things here. First, the condition of $f \ne 0$ should hold. Second, the numerical value $c_1$ of the integration constant 
fixes the absolute magnitude of the temperature.  
 
The final physical field quantity which has to be determined is the velocity shape function and distribution. 
Note, that due to our original Ansatz the two velocity components cannot be determined separately from 
each other, only the $g+h$ is possible to evaluate. This can be easily done from (\ref{comp1})
if we introduce the variable $L:= g+h$. 
Now the ODE is 
\eq
L'f + Lf' - \frac{\eta f'}{2} = 0. 
\eqe
The formal solution now became trivial, namely 
\eq
L = g + h = \frac{\int_0^{\eta} \omega f(\omega) d\omega	+ c_2 }{2f(\eta)}. 
\label{v_compr}
\eqe
This means that our Ansatz is not unique for the velocity field because the $x$ and $y$ coordinates are handled on the same footing. 
The in-depth numerical analysis of  the density (\ref{f_dens_compr}) 
and the velocity (\ref{v_compr}) shape functions lies outside the scope of the present study. 
 
Here, we just wanted to present that incompressible and compressible flow systems having initially comparable PDE systems, which describe similar processes, but behave completely differently during a 
self-similar analysis. Such derivations always give a glimpse into the deep mathematical layers of non-linear PDE systems.  
\section{Summary and Outlook}
We analyzed the incompressible and compressible time-dependent boundary flow equations with additional heat conduction mechanism with the self-similar Ansatz. Analytic solutions were derived for the incompressible flow. The velocity fields can be expressed with the error functions (in some special cases with Gaussian functions) and the temperature with the Kummer functions. The last one has the most complex mathematical structure including some oscillations. 

 It is often asked what are analytic results are good for, we may 
say that our analytic solution could help to test complex numerical fluid dynamics program packages, new numerical routines \cite{endre} or PDE solvers.
For a $t = t_0$ starting time point the time propagation is exactly given by the analitic formula and can be compared to the results of any numerical scheme. 
 
In the second part of our treatise we investigated the compressible  time-dependent boundary flow equations with additional 
heat conduction again with the self-similar Ansatz. For closing constitutive equation, the ideal gas EOS was used.  
It is impossible to derive analytic solutions for the dynamical variables from the coupled ODE system. However, 
highly non-linear independent ODEs exist for each dynamical variables which can be integrated numerically.  An in-depth analysis could be the subject of a next publication.  Work is in progress to apply our self-similar method to  more realistic complex boundary layer flows containing viscous heating or other mechanisms.    
\section{Authors Contributions}
The corresponding author (Imre Ferenc Barna) had the original idea of the study, performed all the calculations, created the figures and wrote 
large part of the manuscript. 
The second and third authors  (Kriszti\'an Hricz\'o and Gabriella Bogn\'ar) checked the written manuscript, improved the language of the final text and 
gave some general instructions. 
The third author (Gabriella Bogn\'ar) organized the financial support and the general founding. 
The last author (L\'aszl\'o M\'aty\'as) checked the literature of the investigated scientific field, corrected the manuscript and had an everyday contact with 
the first author.   
\section{Acknowledgments}
One of us (I.F. Barna) was supported by
the NKFIH, the Hungarian National Research
Development and Innovation Office. This study was supported by project no. 129257 implemented with the 
support provided from the National Research, Development and
Innovation Fund of Hungary, financed under the $K \_ 18$ funding scheme.
\section{Conflicts of Interest}
 The authors declare no conflict of interest.
\section{Data Availability}
The data that supports the findings of this study are all available within the article. 
 
\end{document}